\documentclass[aps,prd,twocolumn,superscriptaddress,showpacs,nofootinbib,altaffilletter]{revtex4-1}
\usepackage{amsmath,amssymb}\begin{document}
\title{Comment on ``From geodesics of the multipole
solutions to the perturbed Kepler problem''}
\author{L. Fern\'andez-Jambrina}
\email[]{leonardo.fernandez@upm.es}
\homepage[]{http://dcain.etsin.upm.es/ilfj.htm}
\affiliation{Matem\'atica Aplicada, E.T.S.I. Navales, Universidad
Polit\'ecnica de Madrid,\\
Arco de la Victoria s/n, \\ E-28040 Madrid, Spain}%
\date{\today}
\begin{abstract}
In this comment we explain the discrepancies mentioned by
the authors between their results and ours about the influence of
the gravitational quadrupole moment in the perturbative calculation of
corrections to the precession of the periastron of quasielliptical
Keplerian equatorial orbits around a point mass.  The discrepancy
appears to be consequence of two different calculations of the angular momentum
of the orbits.
\end{abstract}

\maketitle

In \cite{binet} the authors make use of their static and axisymmetric
solution of Einstein's vacuum equations with a finite number of
multipole moments \cite{geroch,hansen,cornelius} in a system of
coordinates adapted to multipole symmetry to derive a Binet equation
for orbits in the equatorial plane and relate it to the classical
Keplerian problem.  This allows them to write down the relativistic
corrections to Newtonian elliptical orbits in terms of multipole
moments of the source of the gravitational field,
\begin{equation}
 \frac{d^2u}{d\varphi^2}+u=\frac{M}{J^2}+ 3M
u^2-\frac{1}{J^2}\frac{d}{du}V_p^{RMM}(u),
\label{binetclagen}
\end{equation}
where $u$ is the inverse of the radial coordinate, $M$ is the central
mass, $J$ is the angular momentum of the orbit and $V_p^{RMM}(u)$ is a
generalized gravitational potential which encloses the perturbations
due to the gravitational quadrupole moment $Q$ after substracting the
Schwarzschild and centrifugal terms.  We have taken the gravitational
constant $G$ and the mass of the orbiting test particle $m$ equal to one.

The authors obtain a result for the angle precessed by the periastron
in a revolution around the quasielliptical orbit,
\begin{equation}
\Delta\phi=6\pi\left\{\zeta+\zeta^2 
\frac{Q}{M^3}\left(-\frac{1}{2}+3\frac{M}{a}\right)\right\},
\label{precess}
\end{equation}
where $a$ is the semi-major axis of the unperturbed orbital ellipse of
eccentricity $e$ and, hence, of angular momentum $J=\pm\sqrt{a(1-e^2)M}$
and energy $E=-M/2a$.  A dimensionless small parameter
$\zeta=M^{2}/J^2$ has been introduced.

In \cite{multipole} a different approach was followed to a similar 
purpose. Starting with the general metric for a stationary axially 
symmetric vacuum spacetime,
\begin{equation}
ds^2=-f(dt-Ad\phi)^2+\frac{e^{2\gamma}(dr^2+r^2\,d\theta^2)+r^2\sin^2\theta\,d\phi^2}{f},
\end{equation}
where $t$ and $\phi$ are the coordinates associated with the
isometries of the spacetime and the functions $f$, $A$ and $\gamma$
depend only on the coordinates $r$ and $\theta$, a Binet equation is 
written for $U=1/r$ along timelike geodesics in the spacetime,

\begin{equation}
{U_\phi}^2=e^{-2\gamma}\left\{\frac{E^2-f}{f^2\,(J-E\,A)^2}-U^2\right\},
\label{binet1} \end{equation}
where $E$ and $J$, respectively energy and angular momentum per unit
of mass, are the conserved quantities of geodesic motion asociated to
the isometries of the spacetime, \begin{equation} E=f\,(\dot
t-A\dot\phi), \quad J=f\,A\,(\dot
t-A\dot\phi)+\frac{r^2}{f}\,\dot\phi, \end{equation} and the dot
means derivation with respect to proper time along the geodesic.

This Binet equation arises from the normalization condition of the
velocity, $v=(\dot t,\dot r, \dot\theta,\dot \phi)$, of geodesics
parametrised by proper time, $v\cdot v=-1$,
\begin{equation} -1=-f(\dot t-A\dot\phi)^2+\frac{e^{2\gamma}\,\dot
r^2+r^2\,\dot\phi^2}{f},\label{timelike} \end{equation}
and imposing the existence of conserved quantities in order to remove 
the derivatives $\dot\phi$, $\dot t$. Binet's equation (\ref{binet1}) 
is obtained dividing by $\dot\phi^2$ and thereby eliminating the dependence 
on proper time.

In order to compare our results with \cite{binet}, we take
$A(r,\theta)=0$ in order to consider only static spacetimes.

This equation is solved perturbatively \cite{multipole} in powers of a small
dimensionless parameter $\epsilon=M/J$ and a change of variable 
$\psi=\omega\phi$, which allows us to get rid of secular terms in the 
perturbation scheme. This frequence $\omega$ is responsible for the 
precessed angle,
\begin{eqnarray} \label{precess1}
    \frac{{\Delta\phi}}{\pi}&=&\frac{1}{\omega}-1\simeq 6\epsilon^{2}
+\left (\frac{105}{2}+15E_0-\frac {3Q}{M^3}\right )\epsilon^{4},
\end{eqnarray}
whereas the energy is also expanded
\begin{equation} E\simeq1+E_{0}\epsilon^2+\left(-6-10E_0-
\frac{E_0^2}{2}\right)\epsilon^4, \end{equation}
in powers of $\epsilon$. The term $E_{0}$ is related therefore to the 
classical orbit.

The discrepancy mentioned in \cite{binet} arises on identifying the 
small parameters in both expansions by means of $\zeta=\epsilon^2$. 
The term $3M/a$ in (\ref{precess}) appears to be missing in 
(\ref{precess1}).

The explanation is simple and is due precisely to the previous
identification.  The authors assign the classical Keplerian values
$E_{c}=-M/2a$, $J_{c}^2=a(1-e^2)M$ to the energy and momentum of the
elliptical orbit.  However, in their calculations $E$ and $J$ are also
the conserved quantities of geodesic motion and hence $J$ has the
same meaning in both notations, though the value $J^2\simeq a(1-e^2)M$
should enclose the contribution of the quadrupolar moment.

Since we are interested in the first correction to the angular 
momentum, we perform just the classical calculation. 

In classical mechanics the orbit of a particle moving under a central
force of potential $V(r)$ has two conserved quantities, the angular
momentum $J$ and the energy per unit of mass $E$,
\[J=r^2\dot\phi,\quad E=\frac{\dot
r^{2}}{2}+\frac{r^2\dot\phi^2}{2}+V(r)=\frac{\dot
r^{2}}{2}+\frac{J^2}{2r^2}+V(r),\]
in spherical coordinates $(r,\theta,\phi)$.

The gravitational potential due to a central mass $M$ and a 
quadrupole $M$ is not central,
\[V(r,\theta)=-\frac{M}{r}+\frac{Q(3\cos^2\theta-1)}{r^{3}},\]
but considering just orbits in the equatorial plane, it acts as if it 
were central with $V(r)=-M/r-Q/r^3$.

For simplicity we consider circular orbits of radius $r=a$. It is 
clear that they cannot be used for measuring precessed angles, but 
still they provide an easy computation of (\ref{precess}) and 
(\ref{precess1}). They are located at extrema of the effective 
potential,
\[V_{eff}(r)=\frac{J^2}{2r^2}-\frac{M}{r}-\frac{Q}{r^{3}},\]
and so their radius is a solution of
\[Ma^2-J^2a+3Q=0.\]
For $J^4>12MQ$ there are two circular orbits, but we are interested 
in the exterior one,
\[a=\frac{J^2+\sqrt{J^4-12MQ}}{2M}\simeq 
\frac{J^2}{M}-\frac{3Q}{J^2},\]
since it is the one that appears as a perturbation of the classical 
orbit for small $Q$.

The conserved quantities por these circular orbits,
\[E=\frac{J^2}{2a^2}-\frac{M}{a}-\frac{Q}{a^{3}},\quad
J^2=Ma+\frac{3Q}{a},\]
are seen to have simple classical corrections due to the presence of 
the quadrupole moment.

In fact, if in (\ref{precess}) we include the quadrupolar 
correction to the Keplerian angular momentum, $J^2=J_{c}^2+3Q/a$, the 
lowest order, the Schwarzschild term,
\[\frac{6\pi M^2}{J^2}\simeq \frac{6\pi 
M^2}{J_{c}^{2}}\left(1-\frac{3Q}{J_{c}^2a}\right)=
6\pi \left(\zeta-\frac{3Q}{M^2a}\zeta^2\right),\]
a counterterm appears that cancels out the last term in (\ref{precess}).

Hence we have shown that the apparent discrepancy between the formulae
for the precession of the periastron of an equatorial orbit around a
mass endowed with quadrupole moment calculated in \cite{binet} and 
\cite{multipole} is solved by including a first order classical 
contribution to the Keplerian angular momentum due to the 
gravitational quadrupole moment in \cite{binet}.

\end{document}